# Multi-Mode Deep Strong Coupling in a Multi Quantum Well Fabry-Perot Cavity

Lucy Hale[1,*], Johan Andberger[1], Ethan Koskas[1], Frieder Lindel[2,3], Mattias Beck[1], Giacomo Scalari[1], Jérôme Faist[1]

[1]ETH Zürich, Institute of Quantum Electronics, Auguste-Piccard-Hof 1, 8093 Zürich, Switzerland

[2]ETH Zürich, Institute for Theoretical Physics, Zürich 8093, Switzerland

[3]ETH Zürich, Quantum Center, Zürich 8093, Switzerland

*corresponding author email: luhale@phys.ethz.ch

## Abstract

**We present multi-mode deep-strong coupling in a multi-quantum well ($N = 166$) heterostructure. The heterostructure itself acts as a Fabry-Perot cavity, for which the even cavity modes strongly couple to the cyclotron resonance to form Landau polaritons. The experimentally observed vacuum Rabi splitting is larger than the mode spacing and well into the deep-strong coupling regime ($\eta > 1$) resulting in a rich multi-mode polaritonic spectrum which is accurately reproduced by an "all-to-all" multi-photonic, multi-electronic Hopfield coupling model. Remarkably, light-matter decoupling is observed across the whole measurable spectrum, including in the low frequency limit ($\lambda \gg L_{\mathrm{cav}}$) where the normalized coupling strength reaches $\eta = 8.1$. The system demonstrates a robust platform for exploring extreme coupling regimes and its chiral nature holds potential for chiral cavity and chiral mirror applications.**

## Introduction

When light is strongly coupled to matter, the rate of energy exchange between photonic and material components surpasses the dissipative losses in the system, leading to the formation of hybridised light–matter states known as polaritons. Increasing the coupling strength has led to the exploration of the ultrastrong[1-9] and deep strong coupling[10-12] regimes of light-matter interaction, where the coupling strength constitutes a significant fraction of, or even exceeds, the resonance frequencies of the system. These regimes have enabled the observation of phenomena such as vacuum-field-modified electron transport[13,14], enhanced optical nonlinearities[15], polariton–polariton scattering[16], and light-matter decoupling[10]. Other profound modifications of material ground states have been theoretically predicted to arise from ultrastrong coupling to cavity vacuum fields even in the absence of external light[17,18], including cavity-controlled phase transitions[19-23] and emission of squeezed photon pairs from the ground-state[24,25]. Further increasing the strength of light-matter coupling will be highly beneficial for the experimental verification of these phenomena.

Landau polaritons - formed by coupling cavity photons to the cyclotron resonance in a two-dimensional electron gas - have proven to be a very successful approach for realizing the ultrastrong coupling regime [5,10,11,26] . One version of this experiment employs terahertz (THz) frequency metallic metasurfaces as cavities, where coupling is maximized via deeply subwavelength light confinement. This approach has enabled record coupling strengths, sometimes so large that the coupling spans multiple photonic modes [10,11]. However, the metasurface geometry typically supports only linearly polarized modes, preventing access to the inherent chirality of Landau polariton systems that results from broken time-reversal symmetry due to electron cyclotron motion. Although metasurfaces supporting circularly polarized modes are possible[27], preventing mixing between modes of opposing handedness remains challenging. Alternatively Fabry-Perot cavities can be used, which typically exhibit lower coupling strengths but allow for high cooperativity and support circularly polarized fields, enabling the investigation of chiral effects such as the Bloch-Siegert shift [6, 28-31].

Here, we investigate light–matter coupling in a simple inverted Fabry-Perot cavity which consists of a dielectric slab containing a multi-quantum well heterostructure positioned at the center (Fig. 1a). In prior studies of Landau polaritons in Fabry–Perot cavities, metallic or Bragg mirrors have been employed, whose boundary conditions insist on electric field nodes at the cavity edges. This results in a fundamental mode with maximum electric field at the center, enabling efficient coupling to electrons in quantum wells located centrally. In contrast, for the dielectric slab used in this work, the mirror boundaries are formed by Fresnel reflection at the heterostructure-vacuum interface, and therefore the cavity supports a fundamental mode with a field node at the cavity center and antinodes near the cavity ends. Therefore, it is the even-order modes that couple efficiently to the electron cyclotron resonance. In particular, this configuration also supports a low-frequency zeroth ($0^{\mathrm{th}}$) order mode when the cavity thickness $d \ll \lambda$ which is characterized by a uniform, non-zero field distribution across the cavity length. Although this mode was observed in prior studies[32,33], it is still an open question as to how efficiently matter can couple to it. In this work we demonstrate light-matter coupling for the $0^{\mathrm{th}}$ order mode with strength many times larger than the bare cavity frequency, enabled by the simple, compact cavity with deeply sub-wavelength dimensions which also conserves in-plane momentum.

The large coupling and inherent chiral nature of this system suggests possible use as chiral mirrors in future cavity architectures. When strongly coupled to external matter excitations, these structures could be used to imprint chiral characteristics onto material responses, with potential implications for symmetry-breaking and topological effects in light–matter systems[27,34-39].

## Results and Discussion

### Multi-Quantum Well Fabry Perot Cavity

A schematic of the 136 μm-thick cavity with 166 quantum wells in the center is shown in Figure 1a (inset). The computed transmission of the 'empty cavity' without free electron density is also shown in Fig. 1a (blue line). This displays a classic Fabry-Perot spectrum with equally spaced transmission peaks corresponding to the resonant mode frequencies. The first transmission maxima is at zero frequency and is the so-called $0^{\text{th}}$ order mode of the Fabry-Perot cavity, when the cavity thickness $d \ll \lambda$[32]. In addition, using a macroscopic QED treatment of the setup[40] we also calculate the spectral density of photonic states $J(\omega, z)$ (red line in Fig. 1a) that fully determines the coupling of the electrons to the electromagnetic cavity field at zero in-plane wave-vector (see Supplementary Material). The Lorentzian peaks in the spectral density correspond to the photonic modes of the inverted Fabry-Perot cavity [41], whose peak position determines the resonance frequency. In addition, the area under the peak determines how strongly electrons can couple to the photonic mode. In contrast to the empty cavity transmission spectrum, in the spectral density, the $0^{\text{th}}$ order mode peaks at a finite frequency, $\omega_{\text{cav}}^{0}$, showing that the $0^{\text{th}}$ order mode is a finite-frequency resonance to which electrons in the 2DEG can couple. This has a profound effect on the coupling strength, as shown later.

The electric field distribution across the cavity length for the first few cavity modes of the empty cavity are shown in Fig. 1b). Each of these modes corresponds to a peak in the transmission spectrum in Fig. 1a. The modes have alternating odd and even symmetries, with even modes having near-maximum magnitude in the center of the cavity. This includes the $0^{\text{th}}$ order mode, where the cavity supports a constant non-zero field amplitude, meaning that electrons located anywhere inside the cavity can couple to the mode with the same coupling strength. In contrast, the odd modes have near-zero field at the position of the quantum wells.

When a magnetic field is applied, the cavity modes couple to the cyclotron resonance in the 2DEG. A perpendicular magnetic field quantizes the collective electron cyclotron motion into Landau levels. The inter-Landau level transition resonant frequency is given by $\omega_c = eB/m^*$ where $m^* = 0.069m_e$ is the effective electron mass, related to the free electron mass, $m_e$. The 2DEG in the magnetic field therefore exhibits an anisotropic, gyrotropic permitivitty where the in-plane permittivity is determined by the electron cyclotron resonance. The system can be fully modeled using a transfer matrix formalism for circularly polarized light, where the GaAs regions are a dielectric isotropic media and the quantum well region is an effective gyrotropic medium whose resonant frequency depends on magnetic field. The electric field is calculated by multiplying transfer matrices at the interfaces and propagation matrices within the layers. More information about the transfer matrix calculations can be found in the Supplemental Information.

Due to the large number of electrons the cyclotron resonant frequency is superradiantly broadened[42], resulting in efficient coupling of the lossy, broadened cavity modes of the dielectric slab to the cyclotron resonance. The extent of the coupling between the cyclotron resonance and the photonic modes of the cavity is determined by the overlap of the photonic mode with the central region containing the quantum well heterostructure (gray area in Fig. 1b). The symmetry of the modes therefore imposes alternating high-low coupling with odd-even modes.

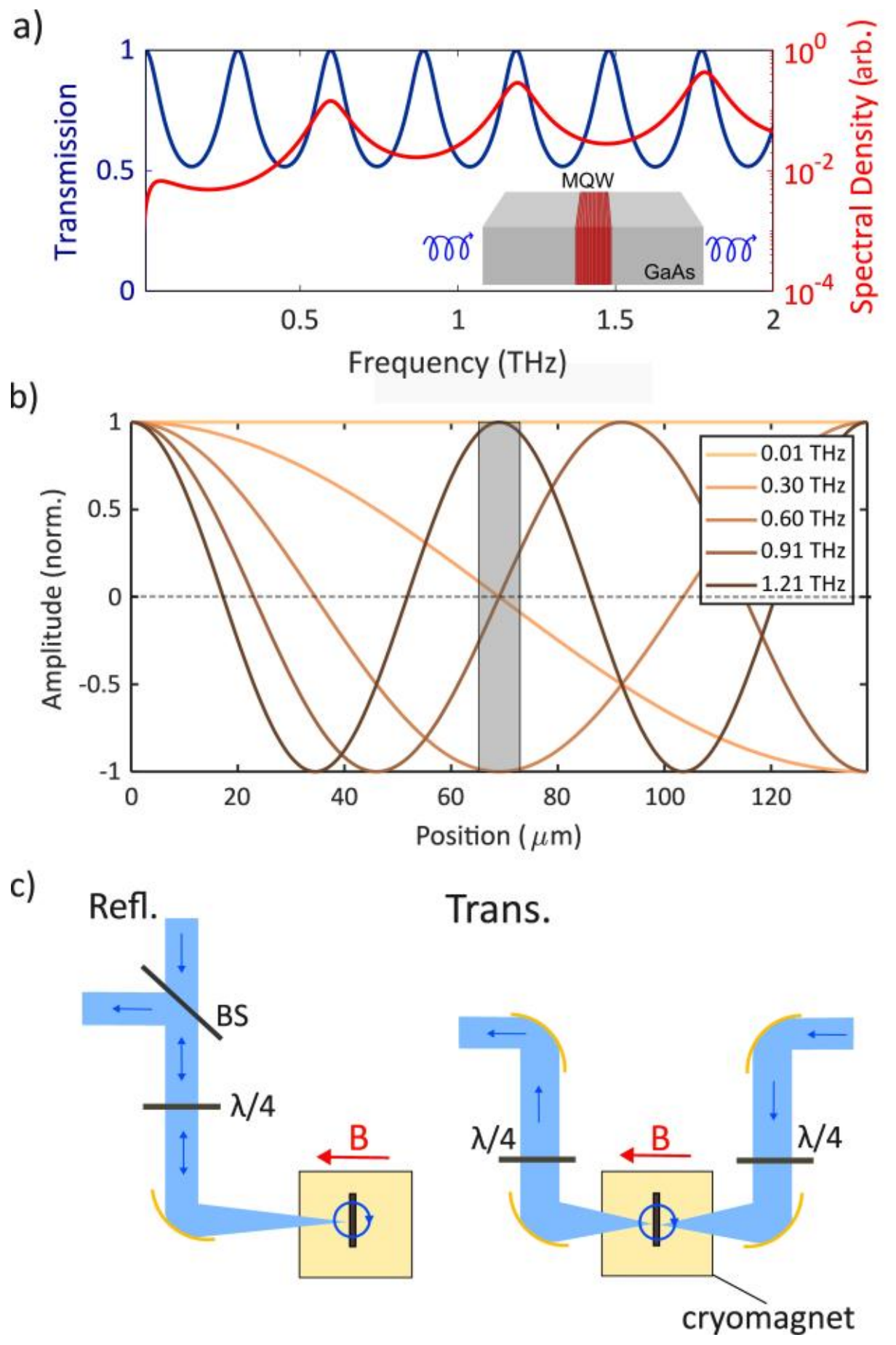

**Fig. 1** Multi Quantum-Well Cavity. a) Calculated transmission (blue) of inverted Fabry-Perot cavity for zero electron density (equivalent to no quantum wells, $n_{2D} = 0$) and calculated cavity spectral density (red). Inset: Schematic of cavity with multi-quantum well (MQW) heterostructure at the center, excited with circularly polarized THz light. b) Analytically calculated mode profiles in the Fabry-Perot cavity for zero electron density. The gray area displays the quantum well region. c) Schematic of the reflection and transmission measurement configurations including beam splitter (BS) and quarter ($\lambda/4$) wave-plates. In both cases the magnetic field is applied perpendicularly - along the QW growth direction.

## Terahertz Magneto-Spectroscopy

Terahertz time-domain magnetospectroscopy measurements of the structure were performed in both transmission and reflection configurations. The experimental set-up is shown in Fig. 1c. The sample is cooled to 3K and a tunable magnetic field is applied perpendicular to the sample surface (in the QW growth direction) from -7T to 7T. In the set-up $\lambda/4$ waveplates are used to convert linearly polarised THz light to right-handed circularly polarized light, and convert the transmitted or reflected circularly polarized THz field back to linear polarization for detection in the same polarization. The measured THz spectra, along with the spectra calculated using the transfer matrix model are shown in Figure 2.

In both reflection and transmission (left panel) we observe rich spectral features as a result of the cavity Fabry-Perot modes strongly coupling to the cyclotron across the entire spectral range. The calculated transmission and reflection (right panel) exhibits excellent agreement with the experimentally measured data, with sharper features due to increased losses and broadening of spectral features in the experiment. The spectra is asymmetric about the 0 T field line, owing to the fact that the right-hand circularly polarized light actively couples to the cyclotron resonance only for the positive magnetic field direction. Despite this, at negative magnetic fields we still observe a substantial deviation from the 'empty cavity' spectra, as the even modes are significantly blue-shifted from their uncoupled values. This phenomenon, known as the Bloch-Siegert shift, results from the counter-rotating component of the cavity enhanced vacuum fields, and becomes visible only when modes are ultrastrongly coupled[6]. The observation of this shift therefore already indicates a very strong multi-modal coupling across the whole spectrum.

The spectrum of cavity modes appear to merge and blue shift as the magnetic field becomes positive and approaches the cyclotron frequency. Around the cyclotron frequency (shown by the dashed line) we see a broad band of very low transmission and mostly high reflection reminiscent of a tunable Reststrahlen band, as observed in other multi quantum well structures[43,44]. Above this, several spectral features can be seen merging as they reach the cyclotron frequency from high magnetic field, some tracing a flat line and then sharply red shifting and dramatically narrowing in linewidth.

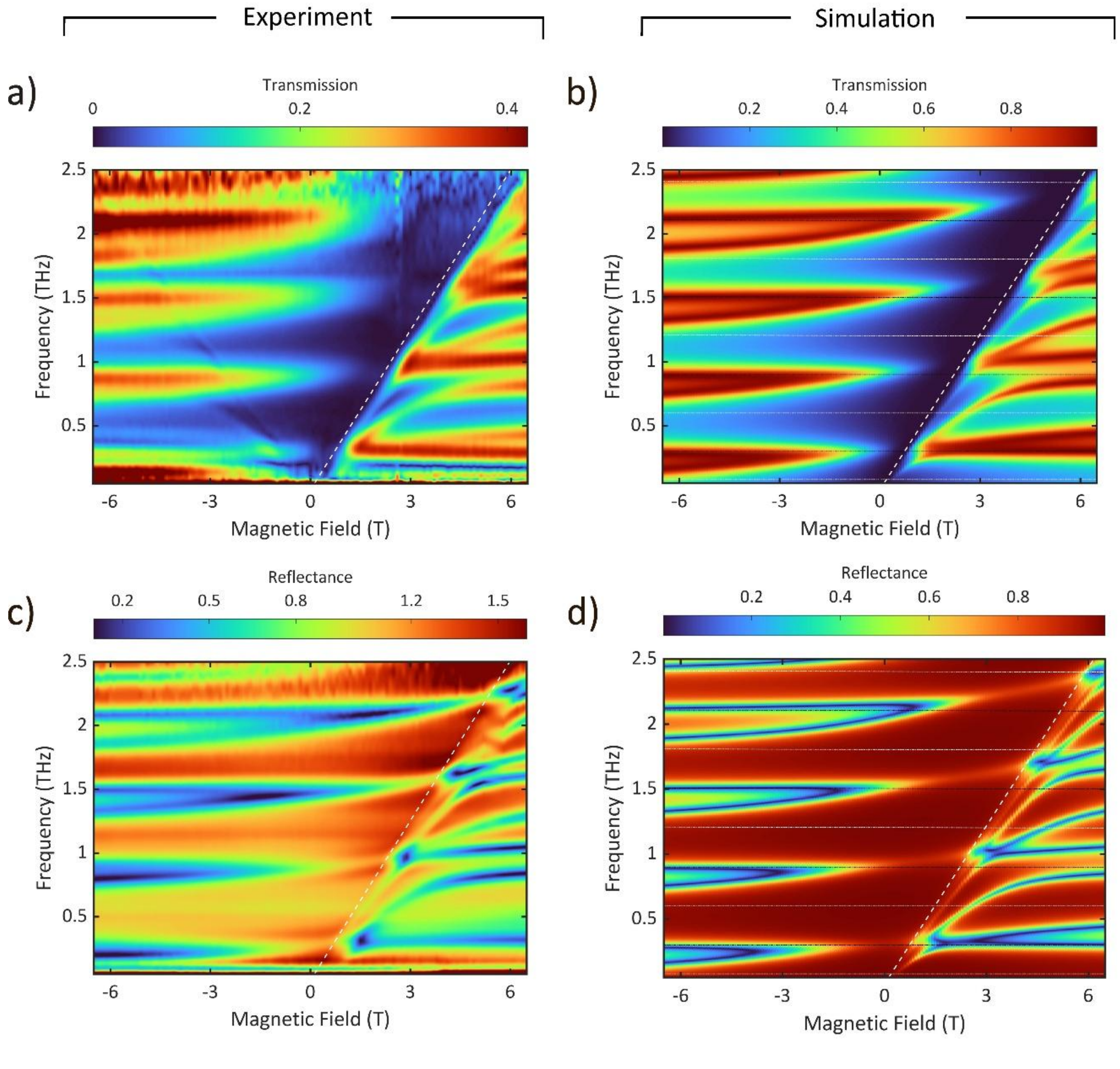

**Fig. 2** Terahertz time-domain magneto-spectroscopy of multi-quantum well Fabry-Perot Cavity. Top: transmission measurement spectra (left) and corresponding spectra calculated using the transfer matrix model (right). Bottom: reflection measurement spectra (left) and corresponding transfer matrix model spectra (right). White dashed line indicates the cyclotron frequency $\omega_c = eB/m^*$. Horizontal dashed lines indicate the position of the empty Fabry-Perot cavity modes (in the absence of electron density), white show modes which have large field overlap with the quantum well region whereas field whereas black shows modes with small overlap.

## Deep Strong Coupling and Decoupling

The presence of many overlapping modes in the spectra makes it difficult to visually interpret the data. Therefore, in order to understand our observations we track how the calculated spectra evolves for increasing electron density (Fig. 3). The coupling strength, indicated by the vacuum Rabi frequency $\Omega_R$, scales with the square root of the electron density as $\Omega_R \propto \sqrt{N_e}$, which in practice scales with the number of quantum wells. Therefore, increasing the electron density in the model allows us to incrementally increase the coupling strength without changing the system geometry.

In Figure 3, for a total sheet electron density corresponding to a single QW (top left panel) we clearly observe that the even Fabry-Perot modes couple strongly to the cyclotron resonance resulting in an anticrossing at the cyclotron frequency, whereas the odd modes do not couple at all. As the density (and thereby effective number of quantum wells) increases, the coupling strength increases and the expanded anticrossing results in an overlap between neighboring odd and even modes, providing an intuitive understanding of the mode pairs seen in the experimental data. As $n_{qw}$ increases further ($n_{qw} > 82$), we see a broad reflectivity band opening up, similar to that observed in the measurement.

The coupling spans multiple modes, so a direct extraction of the global coupling strength is not possible. Similarly, quantifying the normalized coupling strength of each individual mode - defined as half the observed Rabi splitting divided by the uncoupled cavity mode frequency, $\eta_i \equiv \Omega_i/\omega_{cav}^i$ - is also challenging due to the overlap of neighboring coupled modes. Nevertheless, we can obtain an estimate for this value by considering the coupling calculated in the model for a lower number of quantum wells, and using the model to extrapolate to the number of quantum wells in our experiment ($n_{qw} = 166$). For individual even modes labeled $n = 2,4,6$ at 0.6 THz, 1.2 THz and 1.8 THz respectively, we thereby estimate $\eta_2 = 1.14$, $\eta_4 = 0.59$ and $\eta_6 = 0.38$ - demonstrating that the system is well into the deep strong coupling regime for the second mode, and in the ultrastrong coupling regime across the whole measurable spectrum, achieving some of the highest coupling values reported in the literature to date.

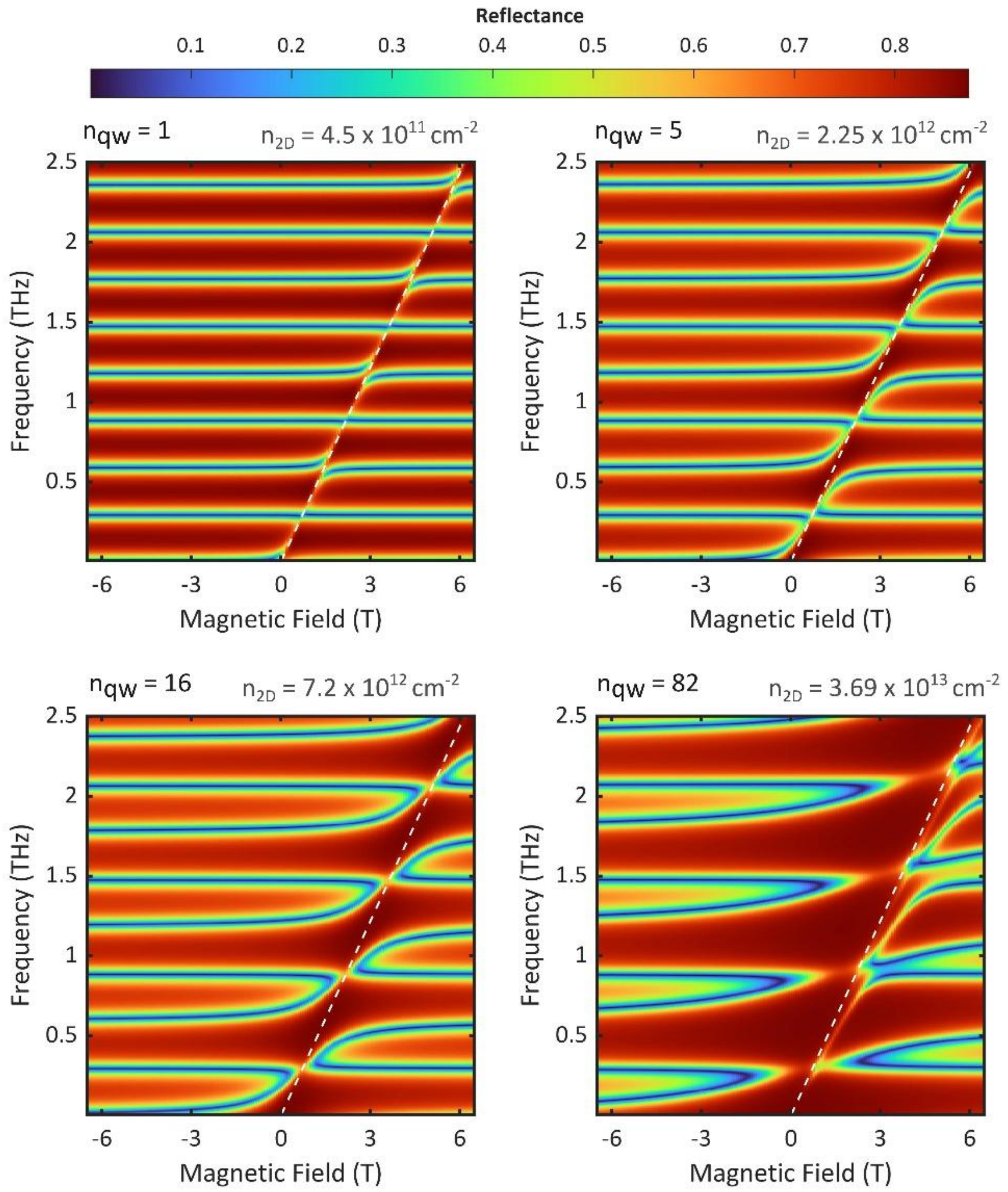

**Fig. 3** Transfer Matrix Modeled Reflectivity for increasing electron density (simulating increasing number of quantum wells with individual electron density of $n_{2D} = 4.5 \times 10^{11} cm^{-2}$). White dashed line indicates the cyclotron frequency $\omega_c = eB/m^*$.

Focusing on the $0^{\text{th}}$ order mode, already for $n_{qw} = 1$ we observe slight coupling of the cyclotron near 0 T, resulting in a “upper polariton-like" feature, resembling one half of an anti-crossing centered at zero frequency. As $n_{qw}$ (and thereby coupling strength) increases, this feature lifts in frequency. The fact that the uncoupled $0^{\text{th}}$ order mode is centered at zero frequency in the transmission spectrum is particularly intriguing as this imposes significant challenges for understanding the coupling and its affect on the ground state. At this point the traditional definition of normalized coupling strength diverges. To evaluate the coupling we therefore rely on the spectral density to redefine the frequency of the mode, which we will discuss later.

We also consider the electric field profile inside the cavity, which can be directly extracted from the transfer matrix model and provides deeper insight into the coupling. Figure 4 shows the electric field profile at different frequencies and magnetic field values (shown by the symbols in Fig. 4a). For the empty cavity ($n_{2D} = 0$, $n_{qw} = 0$) for low frequencies (0.01 THz) with no magnetic field applied, the field in the cavity is close to 1 and constant across the cavity length (black line in Fig. 4b). In contrast, for the 0.3 THz mode which does not couple to the cyclotron resonance (green line in Fig. 4b), the field is at a minimum at the center of the cavity, where the quantum wells are situated. For large $n_{qw}$ as is the situation in Fig. 4, as the low frequency feature shifts up in frequency and approaches the 0.3 THz mode for increasing magnetic field, a minimum of electric field appears at the position of the quantum wells, similar to the 0.3 THz mode. Finally at the point of anti-crossing and the transmission minima (red circle and red line in Fig. 4a,b respectively) the electric field drops to zero at the cavity center. Therefore, counterintuitively, at the anti-crossing the quantum well heterostructure acts as a mirror and the photonic field is expelled, resulting in distinct light matter decoupling. While here this is shown for the low-frequency mode, this can be observed when all higher order modes coincide with the cyclotron frequency across the spectrum. This striking result has been theoretically described[45] and experimentally observed[10] in other Landau polaritonic systems in the deep-strong coupling regime, and represents a 'maxing out' of the light-matter coupling strength, as the coupling becomes so strong that the photonic field is fully reflected, and light and matter effectively decouple.

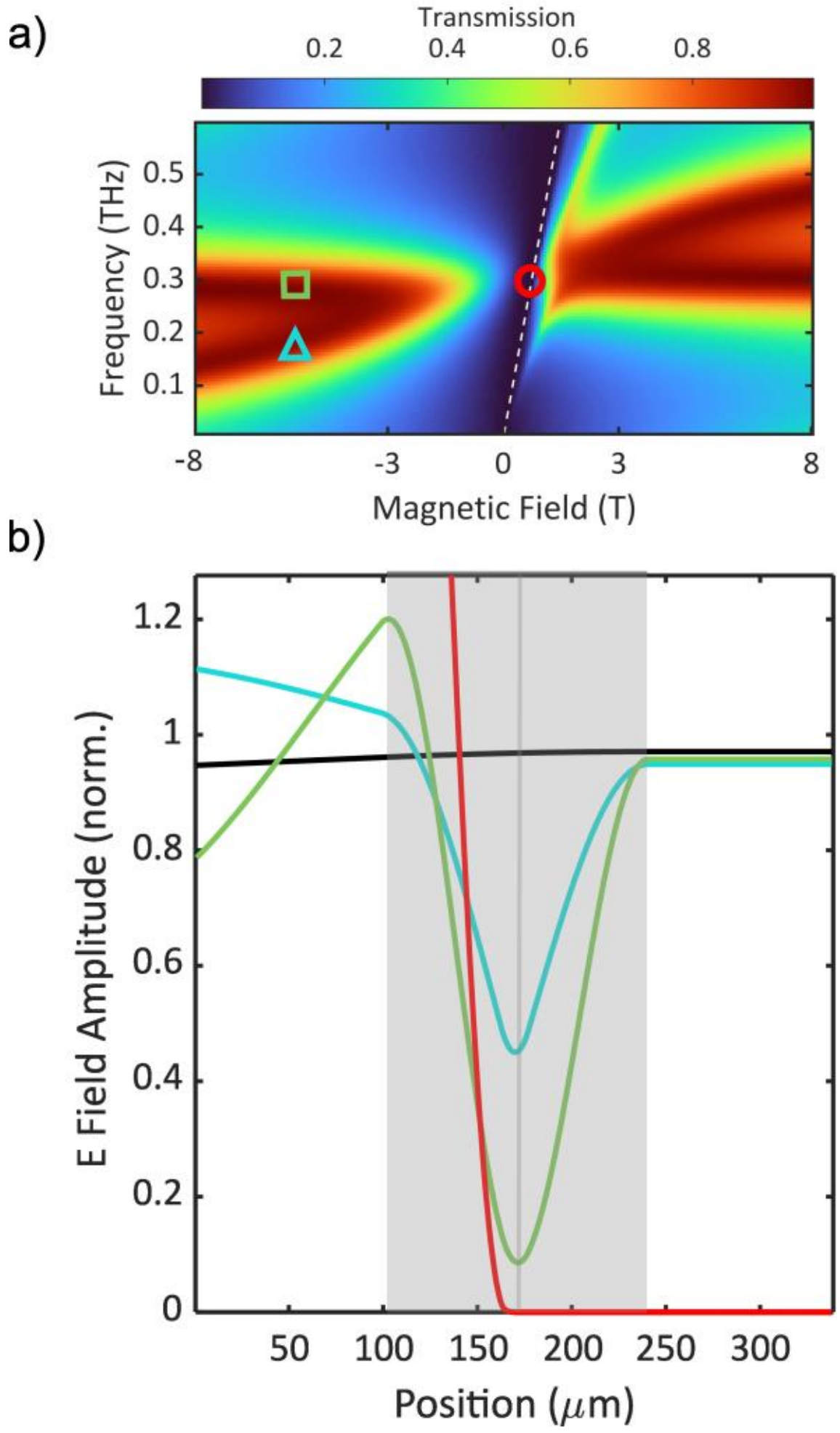

**Fig. 4** a) Transfer Matrix Modeled Transmission for $n_{qw} = 166$. b) Electric field magnitude calculated as sum of forward and backward propagating waves in the transfer matrix calculation (details in the Supplemental Material). Colors correspond to the frequency and B field at the colored symbols in (a). Black line shows the calculated field for an empty cavity ($n_{qw} = 0$) at 0 T and approaching zero frequency (0.01 THz). Grey area shows the dielectric slab thickness and grey line indicates the position of the quantum wells.

## Quantum Mechanical Multimode Model

In addition to the transfer matrix model, we can also obtain an intuitive understanding of the system by calculating the expected polariton frequencies using a quantum mechanical Hopfield model for circularly polarized light developed in Ref. 6. Here, the Hamiltonian of the system is expressed in terms of cyclotron-resonant active and inactive photonic modes, allowing the calculation of both positive magnetic field response (where the cyclotron motion co-rotates with the incident field) and negative magnetic field response (where the cyclotron motion counter-rotates with the incident field). In addition, because the system supports many photonic modes and the multi–quantum-well region spans a non-negligible fraction of their wavelengths, the spatial profiles of the modes must be taken into account. This leads to distinct coupling strengths between each photonic mode and each quantum well, which depend explicitly on the electric-field amplitude at the position of the wells, as shown in Fig. 5a. The second quantized interaction Hamiltonian can then be written as[46]:

$$\mathcal{H}_{\text{int}} = \sum_{n,m} i\,\hbar\Omega_{n,m}\left[b_m^\dagger\left(a_{n,+} + a_{n,-}^\dagger\right) - b_m\left(a_{n,-} + a_{n,+}^\dagger\right)\right], \tag{1}$$

where $\xi$ labels the right $(+)$ and left $(-)$ circular polarization components, $n$ labels the cavity modes, and $a_{n,\xi}^\dagger$ and $a_{n,\xi}$ are the photon creation and annihilation operators for the mode $n$ with polarization $\xi$. The electronic part of the Hamiltonian is expressed in terms of the creation $\left(b_m^\dagger\right)$ and annihilation $(b_m)$ operators of the bright collective bosonic excitation for the inter-Landau level transition. The index $m$ labels the quantum wells, which are all identical except for their spatial position.

$\Omega_{n,m}$ is the coupling between the cavity mode $n$ and the electrons of the quantum well $m$. This

can be expressed as the bare coupling constant for the cavity mode $n$ , $\Omega_n$, multiplied by an overlap factor:

$$\Omega_{n,m} = \Omega_n \alpha_{n,m}. \tag{2}$$

The bare coupling $\Omega_n$ for the mode $n$ can be determined by the electron density $n_{2D}$, cyclotron frequency $\omega_c$, effective electron mass $m_e^*$, vacuum and relative permittivities $\epsilon_0$ and $\epsilon_r$, respectively, cavity length $L_{\text{cav}}$, and mode frequency $\omega_{\text{cav}}^n$ [6]:

$$\Omega_n = \sqrt{\frac{e^2 n_{2D} \omega_c}{m_e^* \epsilon_0 \epsilon_r L_{\text{cav}} \omega_{\text{cav}}^n}}. \tag{3}$$

The overlap factor $\alpha_{n,m}$ quantifies how strong the electric field of mode $n$ is at the position of the quantum well $m$. Figure 5a schematically illustrates the calculation of these overlap factors by showing the normalized electric field profiles of different cavity modes evaluated at the positions of the various quantum wells.

This position-dependent coupling recovers the discrete dipole approximation within each individual quantum well of subwavelength thickness, even when the cavity field varies significantly over the full quantum well stack. As a result, the system exhibits an effective "all-to-all" coupling structure, where each quantum well interacts with each cavity mode with a coupling strength set by the local field amplitude.

By introducing the polaritonic operators and considering their commutator with the full Hamiltonian one can obtain a Hopfield matrix whose eigenvalues correspond to the different polaritonic branches shown as white lines overlayed on the reflectance data in Figure 5b. We find an excellent agreement between the measured reflectance data and our quantum mechanical multimode model. Since the polariton spectrum converges as the $0^{\text{th}}$ order mode frequency $\omega_{\text{cav}}^{0} \to 0$, simulations with sufficiently small $\omega_{\text{cav}}^{0}$ yield coherent results. We take $\omega_{\text{cav}}^{0} = 0.056\,\text{THz}$, which is identified using the spectral density in the next section. More details on the calculation can be found in the Supplemental Material.

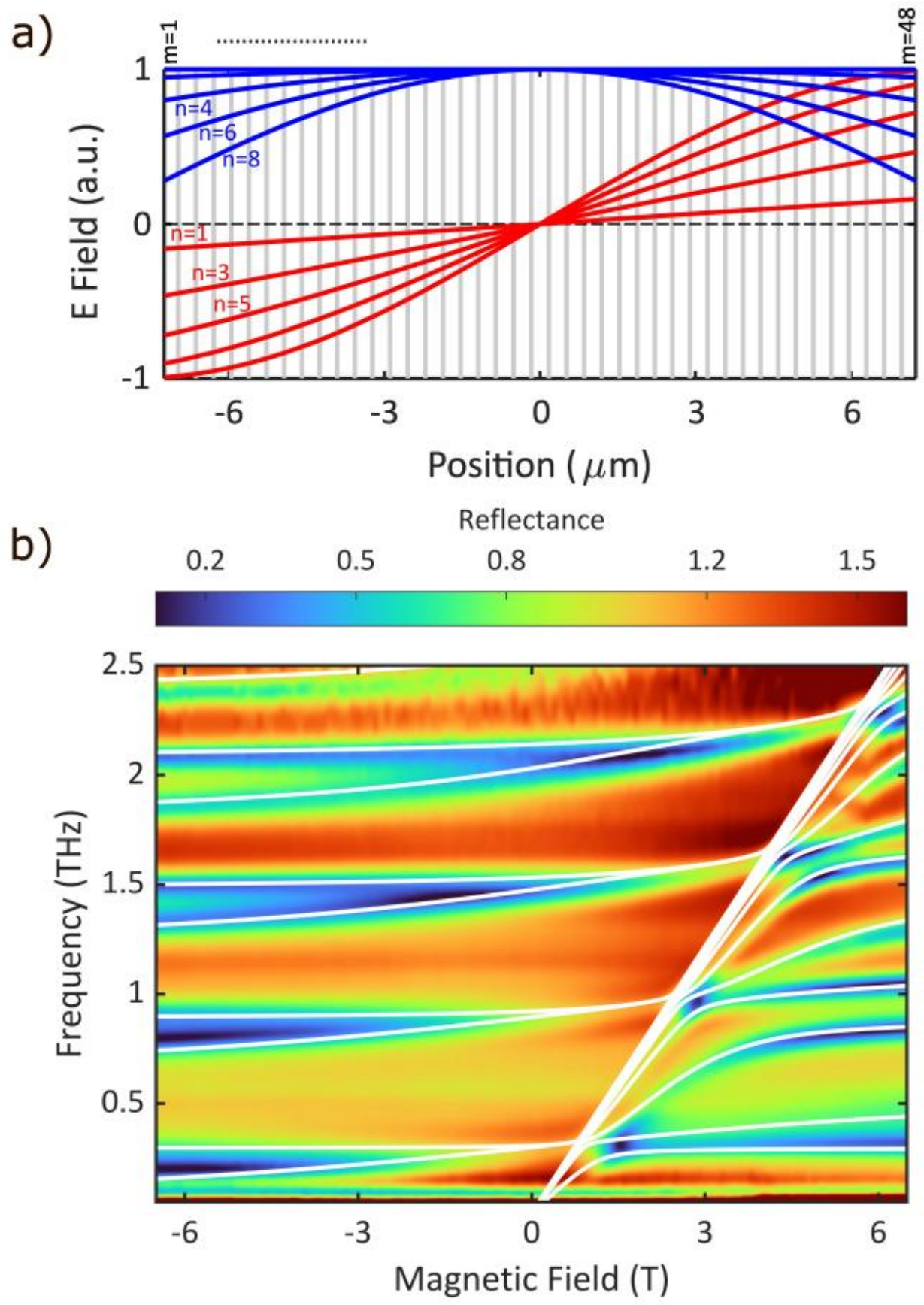

**Fig. 5** a) Schematic showing the field amplitude distribution of individual odd (red) and even (blue) modes across the quantum well region. The individual quantum well regions (labelled 'm') are separated by grey stripes. The overlap of each mode 'n' with the electrons in each region 'm' therefore depends on the spatial position. b) Experimental reflectance spectrum with calculated modes using the Hopfield model (white) overlayed.

Several key features can be inferred from the polariton spectra. Firstly, it is evident that for increasing magnetic fields from -6T to the cyclotron frequency, the odd and even mode pairs merge to create a single feature. Approaching from the high magnetic field side, the sharp curvature of the polariton modes as they reach the cyclotron frequency is a result of the spatially varying photonic field amplitude across the quantum well region. This appears due to the fact that neighboring photonic modes have different field amplitudes in a given region of the quantum well layer and therefore couple differently to each electronic mode. The normalized coupling ratios of all non-zero even modes can be extracted directly from the Hopfield model by calculating $\eta_n = \Omega_n/\omega_{cav}^n$. We obtain $\eta_2 = 1.17$, $\eta_4 = 0.57$ and $\eta_6 = 0.38$, demonstrating very good agreement with the estimations from the transfer matrix model.

## Zeroth Order Mode

As discussed earlier, it is not possible to quantify the coupling strength for the $0^{\text{th}}$ order mode with the standard definition, as the uncoupled cavity mode is centered at 0 THz in the transmission and reflectance spectra. No coupling is strictly possible at zero frequency, due to the vanishing photonic density of states at this frequency.

Therefore, to identify the coupling strength of the $0^{\text{th}}$ order mode, we derive a coupling Hamiltonian in the Supplemental Material based on a Green tensor formalism[40] that takes the full electromagnetic mode continuum into account. From this we extract the spectral density of photonic states $J(\omega, z)$ as shown in 1a. As previously mentioned, the Lorentzian peaks in the spectral density correspond to the photonic modes of the inverted Fabry-Perot cavity[41] , whose peak position and area under the peak determine the resonance frequency and coupling strength (Rabi frequency), respectively. The finite frequency of the peak position of the $0^{\text{th}}$ order mode can be understood as a result of the finite width of the zero frequency mode and the vanishing of the density of states at zero frequency, as has also been recently discussed in the quasi-normal mode framework [33,47].

Analyzing the spectral density, we are therefore able to identify $\omega_{\text{cav}}^0 = 0.058$ THz and the coupling on resonance ($\omega_c = \omega_{\text{cav}}^0$) is given by $\Omega_0 = 0.47$ THz, such that $\eta_0 = 8.1$. More detail on the calculation is given in the Supplemental Information. To our knowledge, this is the highest normalized coupling ratio reported to date, highlighting the strength of using this cavity platform for extreme light–matter interaction. In conventional metamaterial or Fabry–-Perot geometries, much larger cavities would be required to achieve coupling at such low frequencies, resulting in substantially weaker coupling. For example, a normal Fabry–-Perot cavity with a length more than five times greater than in our setup ($L_{\mathrm{cav}} = 720\,\mu\mathrm{m}$) such that its fundamental mode is at $0.058\,\mathrm{THz}$ would have a normalized coupling of only $5.3$ for the same heterostructure according to Eq. (3).

## Conclusion

We have presented a conceptually simple experimental platform for exploring deep strong light–matter coupling using an inverted Fabry Perot cavity formed from a multi-quantum well heterostructure. Terahertz magneto-spectroscopy measurements reveal multimodal deep strong coupling with vacuum Rabi splitting exceeding the cavity mode spacing, resulting in broadband light-matter decoupling which is observed through the expulsion of the photonic field at the anti-crossing point. The system is theoretically understood both classically, through transfer matrix modeling of the spectra and electromagnetic fields, as well as quantum mechanically, by calculating the polariton spectra via the multi-photonic, multi-electronic Hopfield model. Remarkably, the coupling is maximized for the $0^{\mathrm{th}}$ order mode, resulting in a coupling strength almost an order of magnitude larger than the cavity frequency, surpassing even the deep-strong coupling regime.

The cavity's chiral nature suggests that it could be implemented to induce topological effects in material systems through strong light-matter coupling to the enhanced chiral vacuum fields [27,30,34]. Combined with the unique deep strong coupling at very low frequency in a compact device which conserves in-plane momentum, this system could be a platform to observe renormalization of the Von-Klitzing constant in the Quantum-Hall effect[48]. More generally, the self-hybridized multi-quantum well platform is powerful tool for investigating the physics of extreme light–matter interaction.

## Materials and Methods

### Multi-Quantum Well Fabry Perot Cavity Fabrication

The cavity is made by wafer-bonding together two pieces of a 83 GaAs/AlGaAs quantum well heterostructure grown directly on a GaAs substrate. The structure is subsequently lapped equally on either side to form the 136 µm-thick cavity with 166 quantum wells in the center. Each quantum well is a double-side doped square well with an electron density of approximately $n_{2D} = 4.5 \times 10^{11}\ /cm^2$ and a relative effective mass $m^* = 0.069 m_0$. The active region of the quantum wells (excluding the cap layer) is in total 14.5 µm thick. A schematic of the total growth structure is shown in the supplementary material (Fig. S1).

### Terahertz Magnetospectroscopy Measurements

The sample was measured in both reflection and transmission terahertz time-domain spectroscopy (THz-TDS) set-ups. In both cases the sample was at 3K in a superconducting magnet cryostat, enabling measurements in magnetic fields up to 9T. In transmission, a femtosecond pulsed (70 fs) Ti:Sapphire laser was used with a large-area photoconductive antenna (PCA) switch to generate broadband THz pulses. Detection was done with electo-optic sampling in a ZnTe crystal. THz $\lambda/4$ waveplates were used to convert vertically polarized THz light from the PCA switch into circularly polarised light, which was then converted back to linear polarisation with another $\lambda/4$ waveplate before the detection crystal. In reflection, a commerical fiber-based THz-TDS system was used (Menlo Systems). The transmitter and reciever heads were placed at the input and output ports of the beamsplitter shown in Fig. 1 of the main text. In this case the $\lambda/4$ waveplate converts the incoming linearly polarised light to circularly polarised light, then converts it back to linearly polarised light in the orthogonal polarisation upon reflection from the sample. The receiver is therefore oriented perpendicular to the transmitter.

## Transfer Matrix, Hopfield and 0$^{th}$ Order Mode Calculations

Full details of the transfer matrix model calculations, as well as the multimode Hopfield model and its extension to describe the 0$^{th}$ order mode through calculation of the cavity spectral density are given in the Supplementary Information.

## Acknowledgements

The authors would like to acknowledge the Elena Mavrona and Shima Rajabali for work on THz Fabry-Perot cavity strong coupling, as well as Erika Cortese and Simone de Liberato for useful discussions about the multi-mode modeling, and the FIRST-Lab cleanroom where the material growth and processing took place. J.F., G.S., L.H., J.A, E.K acknowledge funding through the SNF grants 207795 and 10000397. This work was supported by the Quantum Center Research Fellowship and the Dr Alfred and Flora Spälti Fonds.

## Author Contributions

Experimental measurements, data acquisition, analysis; L.H. and J.A. Mathematical modeling and analysis; J.A, E.K, F.L. Material growth; M.B. Sample processing; J.A. G.S and J.F. supervision, acquiring of funding. Writing and editing; L.H. J.A. E.K. G.S and J.F.

## Data Availability

All data sets generated and analyzed in this article are available from the corresponding authors upon reasonable request.

## Conflicts of interest

The authors declare no competing interests.

## Additional Information

**Supplementary information** is available.

# Supplementary Material:

# Multi-mode Deep Strong Coupling in a Multi Quantum Well Fabry-Perot Cavity

Lucy Hale[1,*], Johan Andberger[1], Ethan Koskas[1], Frieder Lindel[2,3], Mattias Beck[1], Giacomo Scalari[1] Jérôme Faist[1]

[1]ETH Zürich, Institute of Quantum Electronics, Auguste-Piccard-Hof 1, 8093 Zürich, Switzerland
[2]ETH Zürich, Institute for Theoretical Physics, Zürich 8093, Switzerland
[3]ETH Zürich, Quantum Center, Zürich 8093, Switzerland

*corresponding author email: luhale@phys.ethz.ch

## 1. Multi Quantum Well Heterostructures

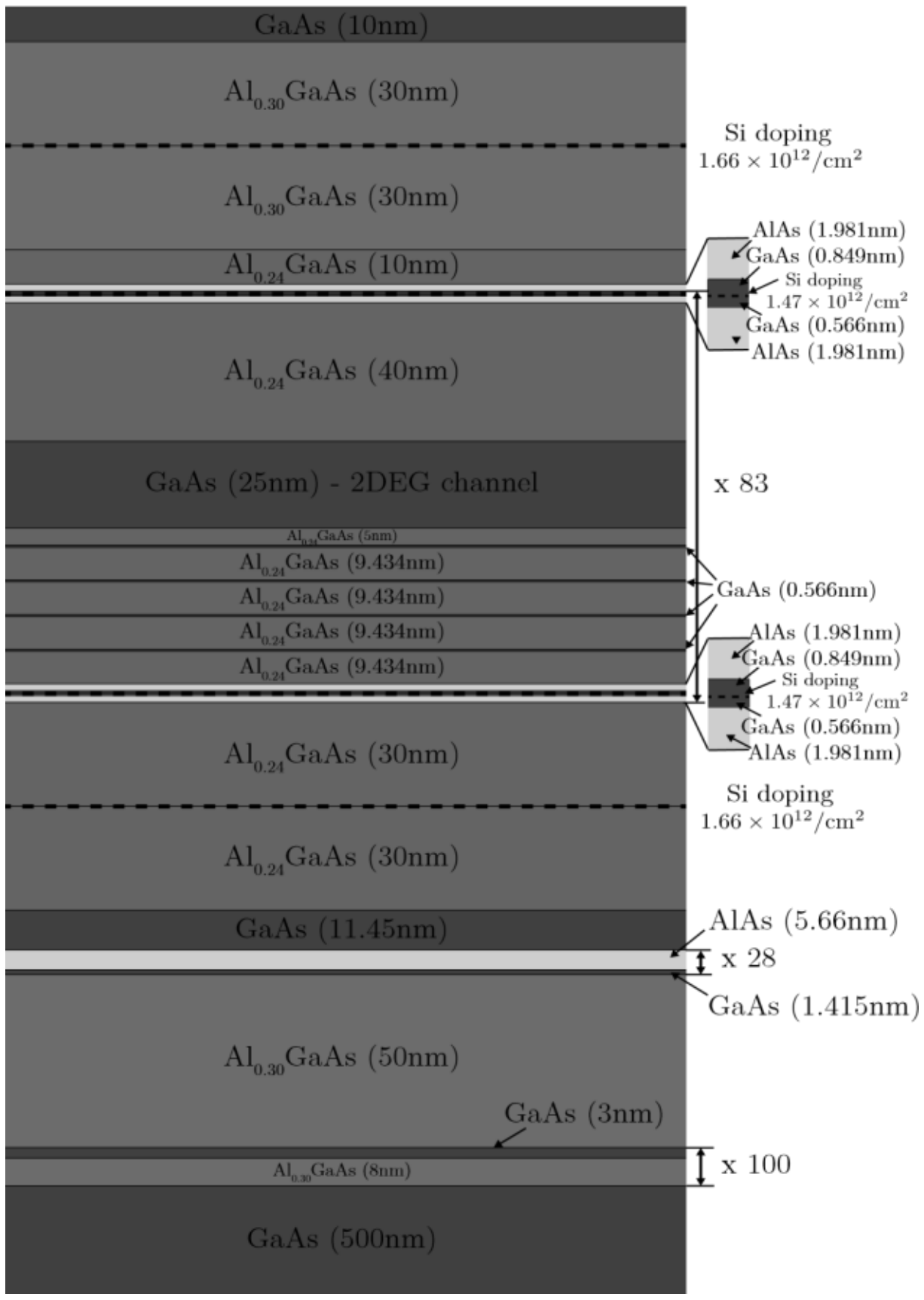

**Fig. S1**. Schematic of the multi-quantum well EV2476 heterostructure used for the fabrication of the samples.

## 2. Transfer Matrix Model

In order to explain the measured spectra we developed a $4 \times 4$ transfer matrix formalism, following the procedure outlined in [1]. Here, the fields at a given position are expressed as a combination of forward and backward propagating waves, and can be written in the basis of either linearly or circularly polarised light (shown in Fig. S2). Here we use circularly polarised light to reflect our measurement conditions.

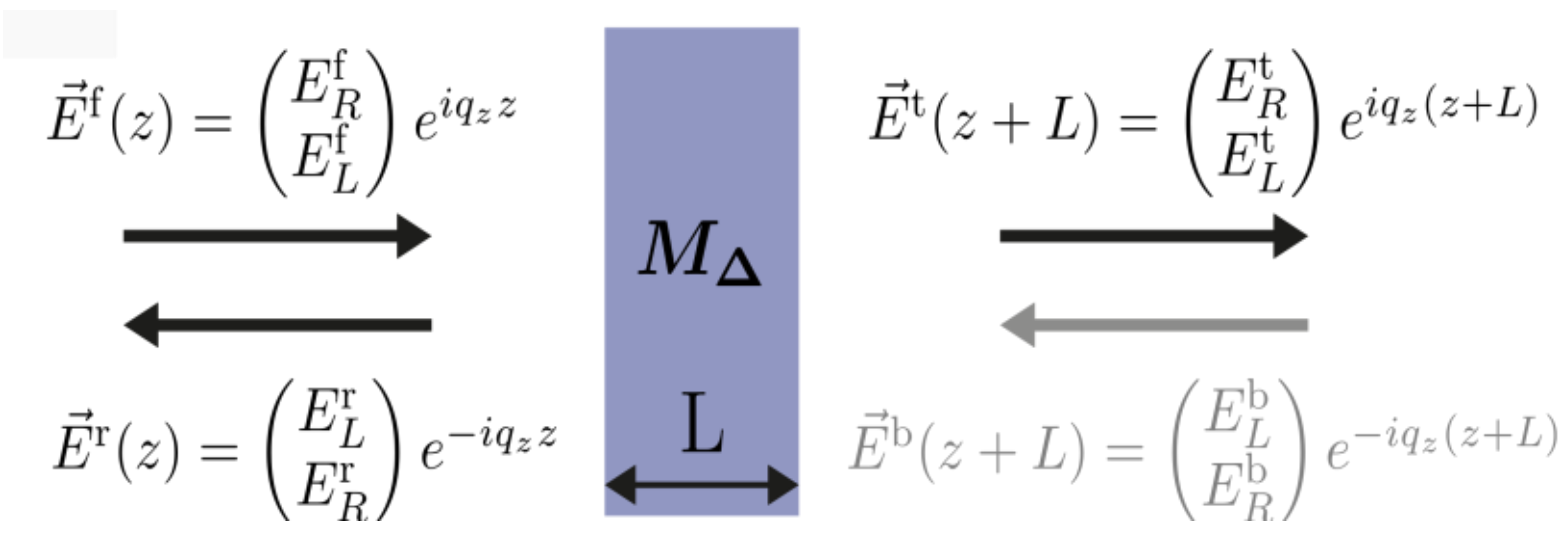

**Fig. S2**. Forwards- and backwards-propagating waves in circular polarization, since the $y$-component is reversed for backwards-propagating waves in linear polarization this means that right-hand circularly polarized light becomes left-hand circularly polarized and vice versa.

The total transfer matrix for the sample is given by the multiplication of transfer matrices in each region:

$$M_T = M_{\Delta_{\mathrm{iso}}}(L_{GaAs}) M_{\Delta_{\mathrm{gyro}}}(L_{QW}) M_{\Delta_{\mathrm{iso}}}(L_{GaAs}) \tag{1}$$

Where $M_{\Delta_{\mathrm{iso}}}$ and $M_{\Delta_{\mathrm{gyro}}}$ are the transfer matrices for an isotropic and gyrotropic medium, respectively, with lengths $L_{GaAs}$ and $L_{QW}$, given by

$$\mathrm{M}_{\Delta_{\mathrm{iso}}}(\mathrm{z}) = \begin{pmatrix} \cos(\mathrm{k_z z}) & 0 & -\mathrm{Z}\sin(\mathrm{k_z\, z}) & 0 \\ 0 & \cos(\mathrm{k_z z}) & 0 & \mathrm{Z}\sin(\mathrm{k_z\, z}) \\ \frac{1}{\mathrm{Z}}\sin(\mathrm{k_z\, z}) & 0 & \cos(\mathrm{k_z z}) & 0 \\ 0 & -\frac{1}{\mathrm{Z}}\sin(\mathrm{k_z\, z}) & 0 & \cos(\mathrm{k_z z}) \end{pmatrix}_{\{\vec{\mathrm{R}},\vec{\mathrm{L}}\}} \tag{2}$$

with $k_z = \pm\frac{\omega}{c_0}\sqrt{\varepsilon_r \mu_r}$, $Z = \sqrt{\frac{\mu_r}{\varepsilon_r}} Z_0$, and

$$\mathbf{M}_{\Delta_{\mathrm{gyro}}}(\mathrm{z}) = \begin{pmatrix} \cos(\mathrm{k_R z}) & 0 & \mathrm{Z_R}\sin(\mathrm{k_R z}) & 0 \\ 0 & \cos(\mathrm{k_L z}) & 0 & -\mathrm{Z_L}\sin(\mathrm{k_L z}) \\ -\frac{1}{\mathrm{Z_R}}\sin(\mathrm{k_R z}) & 0 & \cos(\mathrm{k_R z}) & 0 \\ 0 & \frac{1}{\mathrm{Z_L}}\sin(\mathrm{k_R z}) & 0 & \cos(\mathrm{k_L z}) \end{pmatrix}_{\{\vec{\mathrm{R}},\vec{\mathrm{L}}\}} \tag{3}$$

where

$$k_R = \pm\frac{\omega}{c_0}\sqrt{\frac{(\varepsilon_{11} - i\varepsilon_{12})\mu_r}{2}} = \pm\frac{\omega}{c_0}\sqrt{\varepsilon_R\mu_r}, \tag{4}$$

$$k_L = \pm\frac{\omega}{c_0}\sqrt{\frac{(\varepsilon_{11} + i\varepsilon_{12})\mu_r}{2}} = \pm\frac{\omega}{c_0}\sqrt{\varepsilon_L\mu_r}, \tag{5}$$

and

$$Z_R = \sqrt{\frac{\mu_r}{\varepsilon_R}}Z_0,\ Z_L = \sqrt{\frac{\mu_r}{\varepsilon_L}}Z_0. \tag{6}$$

Here, $R$ and $L$ subscripts represent right and left circular polarisations, which have non-degenerate eigenvalues.

For a gyrotropic material with a certain thickness the permittivity contribution from free charge carriers in a perpendicular magnetic field can be derived in terms of a permittivity tensor that can be written in cartesian coordinates as (see [2]):

$$\varepsilon_r = \begin{pmatrix} \varepsilon_{xx} & \varepsilon_{xy} & 0 \\ -\varepsilon_{xy} & \varepsilon_{xx} & 0 \\ 0 & 0 & \varepsilon_{zz} \end{pmatrix} \tag{7}$$

with

$$\varepsilon_{xx} = \varepsilon_r + \frac{\omega_p^2}{\omega}\frac{\omega + i\gamma}{(\omega + i\gamma)^2 - \omega_{\mathrm{cyc}}^2}, \tag{8}$$

$$\varepsilon_{xy} = \frac{\omega_p^2}{\omega}\frac{i\omega_{\mathrm{cyc}}}{(\omega + i\gamma)^2 - \omega_{\mathrm{cyc}}^2}, \tag{9}$$

$$\varepsilon_{zz} = \varepsilon_r - \frac{\omega_p^2}{\omega(\omega + i\gamma)}, \tag{10}$$

where $\varepsilon_r$ is the relative permittivity of the underlying substrate of the 2DEG, $\gamma$ is the scattering rate (which may be given in terms of a lifetime $\tau = 1/\gamma$) and $\omega_{\text{cyc}} = eB/m^*$ is the cyclotron frequency. The 3D plasma frequency $\omega_p$ is defined as:

$$\omega_p = \sqrt{\frac{n_{3D}e^2}{m^*\varepsilon_0}} = \sqrt{\frac{n_{2D}e^2}{m^*\varepsilon_0 L_{\text{eff}}}}, \tag{11}$$

with $n_{3D}$ being the (effective) 3D electron density, $L_{\text{eff}}$ being the effective length of the medium, and $n_{2D}$ being the 2D electron density for the 2DEGs. Note that $\varepsilon_r$ is absent from the definition of the plasma frequency.

In the basis of circularly polarized light (basis $(R, L)$) the tensor is diagonal $\boldsymbol{\varepsilon}_r = \text{diag}(\varepsilon_R, \varepsilon_L, \varepsilon_{zz})$ with

$$\varepsilon_R(\omega, \omega_{cyc}, \gamma) = \varepsilon_r - \frac{\omega_p^2}{\omega(\omega - \omega_{\text{cyc}} + i\gamma)}, \tag{12}$$

$$\varepsilon_L(\omega, \omega_{cyc}, \gamma) = \varepsilon_r - \frac{\omega_p^2}{\omega(\omega + \omega_{\text{cyc}} + i\gamma)}, \tag{13}$$

since $\varepsilon_R = \varepsilon_{xx} - i\varepsilon_{xy}$, $\varepsilon_L = \varepsilon_{xx} + i\varepsilon_{xy}$. It should be emphasized that $\varepsilon_R$ denotes the permittivity for right-circularly polarized light, while $\varepsilon_L$ denotes the permittivity for left-circularly polarized light.

The relative permittivity of in the terahertz region $< 2$ THz can be approximated as real, the static permittivity is $\varepsilon_{DC} = 12.9$ and the high frequency limit is $\varepsilon_\infty = 10.89$ at $T = 300$ K [3]. All materials can be considered to be essentially non-magnetic and as such $\mu_r = 1.0$.

## 3. Multimode Hopfield Model

We follow Refs. [4]-[6] to describe the interaction between electrons in multiple quantum wells, subject to a perpendicular magnetic field, and the circularly polarised modes of the cavity. Within the Hopfield–Bogoliubov framework, the second-quantized Hamiltonian, as presented in Ref. [4] reads (extended to include multiple quantum wells)

$$\mathcal{H} = \mathcal{H}_{\text{cav}} + \mathcal{H}_{\text{Landau}} + \mathcal{H}_{\text{int}} + \mathcal{H}_{\text{dia}}, \tag{14}$$

where

$$\mathcal{H}_{\text{cav}} = \sum_{\xi=\pm} \sum_n \hbar\, \omega_{\text{cav}}^n \left(a_{n,\xi}^\dagger a_{n,\xi} + \tfrac{1}{2}\right) \tag{15}$$

is the cavity Hamiltonian, $\xi$ labels the right $(+)$ and left $(-)$ circular polarization components, $n$ labels the cavity modes, and $a_{n,\xi}^{\dagger}$ and $a_{n,\xi}$ are the photon creation and annihilation operators for the mode $n$ with polarization $\xi$ and frequency $\omega_{\text{cav}}^{n} = n\pi c / L_{\text{cav}} n_{\text{cav}}$. The *a priori* unknown frequency of the $n = 0$ mode is determined in the next section. However, as mentioned in the main text, the polaritonic spectrum converges when $\omega_{\text{cav}}^{0} \to 0$, so its exact value has a negligible effect on the results provided it is much smaller than the light-matter coupling strength.

$$\mathcal{H}_{\text{Landau}} = \sum_{m=1}^{n_{\text{qw}}} \hbar\,\omega_c \left(b_m^{\dagger} b_m + \tfrac{1}{2}\right) \tag{16}$$

is the electronic Hamiltonian expressed in terms of the creation $\left(b_m^{\dagger}\right)$ and annihilation $(b_m)$ operators of the bright collective bosonic excitation for the inter-Landau level transition with cyclotron frequency $\omega_c$. The index $m$ labels the quantum wells, which are all identical except for their spatial position;

$$\sum_{n,m} \mathrm{i}\,\hbar \Omega_{n,m} \left[ b_m^{\dagger}\left(a_{n,+} + a_{n,-}^{\dagger}\right) - b_m\left(a_{n,-} + a_{n,+}^{\dagger}\right)\right] \tag{17}$$

is the interaction Hamiltonian, which contains the counter-rotating terms. The operators $b_m$ and $b_m^{\dagger}$ couple to matter creation and annihilation operators of opposite circular polarisation. Specifically, the cyclotron resonance couples co-rotatingly to right-circularly polarised photons, while it couples counter-rotatingly to left-circularly polarised photons. The coupling strength $\Omega_{n,m}$ can be expressed as ([4]):

$$\Omega_{n,m} = \sqrt{\frac{e^2 n_{2D} \omega_c}{2 m_e^* \epsilon_0 \epsilon_r \omega_{\text{cav}}^{n}}} f_n(z_m), \tag{18}$$

with $f_n$ the normalized shape of the $n$th mode over the cavity and $z_m$ the position of the $m$th quantum well, the overlap factor $\alpha_{n,m} = f_n(z_m)\sqrt{L_{\text{cav}}/2}$ is defined to be on the order of unity. The interaction Hamiltonian is derived within the dipole (long-wavelength) approximation, applied locally to each quantum well, which is assumed to be sufficiently thin. Nevertheless, the spatial variation of the cavity field across the entire quantum-well stack is explicitly accounted for through the overlap factors.

$$\mathcal{H}_{\text{dia}} = \sum_{n,n',m} \Omega_{n,m}\, \Omega_{n',m} \frac{\hbar}{\omega_c} \left(a_{n,-} + a_{n,+}^{\dagger}\right)\left(a_{n',+} + a_{n',-}^{\dagger}\right) \tag{19}$$

is the diamagnetic Hamiltonian arising from the square modulus of the vector potential.

To write the Hamiltonian in a more compact form and facilitate its diagonalization, we introduce the following matrices and vectors:

$$\boldsymbol{\omega}_{\mathrm{cav}} = \mathrm{diag}(\omega_{\mathrm{cav}}^0, \omega_{\mathrm{cav}}^1, \omega_{\mathrm{cav}}^2, \ldots), \tag{20}$$
$$\mathbf{a}_\pm = \left(a_{0,\pm}, a_{1,\pm}, a_{2,\pm}, \ldots\right)^{\mathrm{T}}, \tag{21}$$

Which have a length determined by the number of cavity modes considered in the model $n_{\mathrm{mode}}$, and

$$\boldsymbol{\omega}_{\mathbf{c}} = \mathrm{diag}(\omega_c, \omega_c, \omega_c, \ldots), \tag{22}$$
$$\mathbf{b} = (b_1, b_2, b_3, \ldots)^{\mathrm{T}}, \tag{23}$$

Which have a length determined by number of quantum wells in the system $n_{\mathrm{qw}}$. This gives:

$$\boldsymbol{\Omega} = (i\Omega_{n,m})_{n,m} \tag{24}$$
$$\mathbf{D} = -\frac{\boldsymbol{\Omega} \times \boldsymbol{\Omega}^{\mathrm{T}}}{\omega_c} \tag{25}$$

of size $n_{\mathrm{mode}} \times n_{\mathrm{qw}}$ and $n_{\mathrm{mode}} \times n_{\mathrm{mode}}$ respectively. The Hamiltonian can then be expressed in the compact form :

$$\begin{aligned}\frac{\mathcal{H}}{\hbar} &= \mathbf{a}_+{}^\dagger \boldsymbol{\omega}_{\mathrm{cav}} \mathbf{a}_+ + \mathbf{a}_-{}^\dagger \boldsymbol{\omega}_{\mathrm{cav}} \mathbf{a}_- + \mathbf{b}^\dagger \boldsymbol{\omega}_{\mathbf{c}} \mathbf{b} \\ &+\mathbf{b}^\dagger \boldsymbol{\Omega}\left(\mathbf{a}_+ + (\mathbf{a}_-{}^\dagger)^{\mathrm{T}}\right) - \mathbf{b}^{\mathrm{T}} \boldsymbol{\Omega}\left(\mathbf{a}_- + (\mathbf{a}_+{}^\dagger)^{\mathrm{T}}\right) \\ &+\left(\mathbf{a}_-{}^{\mathrm{T}} + \mathbf{a}_+{}^\dagger\right)\mathbf{D}\left(\mathbf{a}_+ + (\mathbf{a}_-{}^\dagger)^{\mathrm{T}}\right).\end{aligned} \tag{26}$$

This Hamiltonian can be diagonalized following Ref. [5] by introducing the polaritonic operators

$$\begin{aligned} p_i &= \mathbf{w}_\mathbf{i}^+ \mathbf{a}_+ + \mathbf{w}_\mathbf{i}^- \mathbf{a}_- + \mathbf{x}_\mathbf{i} \mathbf{b} \\ &+\mathbf{y}_\mathbf{i}^+ (\mathbf{a}_+{}^\dagger)^{\mathrm{T}} + \mathbf{y}_\mathbf{i}^- (\mathbf{a}_-{}^\dagger)^{\mathrm{T}} + \mathbf{z}_\mathbf{i} (\mathbf{b}^\dagger)^{\mathrm{T}}, \end{aligned} \tag{27}$$

where we have introduced the Hopfield coefficients as row vectors matching the size of $\mathbf{a}_\pm$ and $\mathbf{b}$. Computing the commutator $[p_i, \mathcal{H}]$ gives the Hopfield block matrix written in the basis of the Hopfield coefficient $\vec{p}_i = (\mathbf{w}_\mathbf{i}^+, \mathbf{w}_\mathbf{i}^-, \mathbf{x}_\mathbf{i}, \mathbf{y}_\mathbf{i}^+, \mathbf{y}_\mathbf{i}^-, \mathbf{z}_\mathbf{i})$:

$$M := \begin{pmatrix} \boldsymbol{\omega}_{\mathrm{cav}} + \mathbf{D} & 0 & -\boldsymbol{\Omega} & 0 & \mathbf{D} & 0 \\ 0 & \boldsymbol{\omega}_{\mathrm{cav}} + \mathbf{D} & 0 & \mathbf{D} & 0 & \boldsymbol{\Omega} \\ \boldsymbol{\Omega} & 0 & \boldsymbol{\omega}_c & 0 & \boldsymbol{\Omega} & 0 \\ 0 & -\mathbf{D} & 0 & -\boldsymbol{\omega}_{\mathrm{cav}} - \mathbf{D} & 0 & -\boldsymbol{\Omega} \\ -\mathbf{D} & 0 & \boldsymbol{\Omega} & 0 & -\boldsymbol{\omega}_{\mathrm{cav}} - \mathbf{D} & 0 \\ 0 & \boldsymbol{\Omega} & 0 & \boldsymbol{\Omega} & 0 & -\boldsymbol{\omega}_c \end{pmatrix}, \tag{28}$$

which satisfies $M\vec{p}_i = \omega_i \vec{p}_i$. The eigenvalues of $M$ give the energies $\hbar\omega_i$ of the different polaritonic branches.

# 4. Macroscopic quantum electrodynamics description

We extend the framework developed in the last section, in which the electromagnetic field is already given in terms of a discrete set of resonant photonic modes, by incorporating the full multimode electromagnetic field into the description. This will allow us to study the mode structure of the inverted Fabry-Perot cavity including the $0^{\text{th}}$-order mode. To this end, we express the field inside the inverted Fabry-Perot resonator in terms of the dyadic Green tensor of the vector Helmholtz equation $\mathbf{G}$ using the framework of macroscopic quantum electrodynamics [7]:

$$\mathbf{E}(\mathbf{r}) = i \int_0^{\infty} \mathrm{d}\,\omega \int \mathrm{d}^3 r' \frac{\omega^2}{c^2} \sqrt{\frac{\hbar}{\pi \epsilon_0} Im\epsilon(\mathbf{r}', \omega)} \\ \times \mathbf{G}(\mathbf{r}, \mathbf{r}', \omega) \cdot \mathbf{f}(\mathbf{r}', \omega) + h.c., \tag{29}$$

where $\mathbf{f}(\mathbf{r}', \omega)$ are bosonic annihilation operators. We treat the Landau levels within the dipole approximation as in Refs. [4-6] by assuming that their polarization is simply given by their quantized center of mass coordinate $\mathbf{X}_i$:

$$\mathbf{P}(\mathbf{r}) = e n_{2\mathrm{D}} \sum_i \mathbf{X}_i \, \delta(z - z_i), \tag{30}$$

$$\mathbf{X}_i = \sqrt{\frac{\hbar}{2Nm^* \omega_c}} \left( \mathbf{e}_x [b_i + b_i^\dagger], + i \mathbf{e}_y [b_i - b_i^\dagger] \right), \tag{31}$$

with $N$ being the total number of electrons, and $b_i$ bosonic annihilation operators. The light-matter coupling Hamiltonian in the multipolar coupling scheme then reads $H_{\mathrm{int}} = \int \mathrm{d}^2 r \mathbf{P}(\mathbf{r}) \cdot \mathbf{E}(\mathbf{r})$. We follow Refs. [8,9] to rewrite $H_{\mathrm{int}}$ in terms of the dipole weighted spectral density and emitter centered bosonic field operators $C_{\sigma,i}(\omega)$:

$$H_{\mathrm{int}} = \int_0^{\infty} \mathrm{d}\,\omega \sum_{i,j=1}^{n_{QW}} \sum_{\sigma = x,y} \frac{1}{\sqrt{2}} W_{\sigma,ij}(\omega) B_\sigma C_{\sigma,j}(\omega). \tag{32}$$

Here, we have defined

$$B_\sigma = \begin{cases} (b + b^\dagger) & \text{if } \sigma = x \\ i(b - b^\dagger) & \text{if } \sigma = y. \end{cases} \tag{33}$$

Furthermore, $(\mathbf{W}_\sigma^\dagger \cdot \mathbf{W}_\sigma)_{ij} = J_{ij}$, where $\mathbf{J}$ is the dipole weighted generalized spectral density that fully determines the light-matter coupling [10] and reads

$$J_{ij}(\omega) = \theta(\omega) \frac{e^2 n_{2\mathrm{D}} \hbar \omega^2}{\epsilon_0 c^2 m_e^* \omega_c \pi} Im\left[G_{xx}\left(\mathbf{k}_\parallel = 0, z_i, z_j, \omega\right)\right]. \tag{34}$$

$G_{\sigma\sigma}(\mathbf{k}_\parallel = 0, z_i, z_j, \omega)$ is $\sigma\sigma$-component of the reciprocal space Green tensor of the vector Helmholtz equation, which is known analytically for the inverted Fabry-Perot cavity under consideration here:

$$\begin{aligned} &G_{xx}\left(\mathbf{k}_\parallel = 0, z_i, z_j, \omega\right) \\ &= \frac{\mathrm{i}}{2k}\left\{\mathrm{e}^{\mathrm{i}k|z_i - z_j|} + \frac{r_s}{D_s}\left[2 r_s \mathrm{e}^{2\mathrm{i}kL_\mathrm{cav}} cos\left[k\left(z_i - z_j\right)\right]\right.\right. \\ &\left.\left. +2\mathrm{e}^{\mathrm{i}kL_\mathrm{cav}} cos\left[k\left(z_i + z_j - L_\mathrm{cav}\right)\right]\right]\right\}, \end{aligned} \tag{35}$$

with $k = \overline{n}\omega/c$, where we define the average refractive index insight the cavity $\overline{n} = (L_{QW}\sqrt{11.31} + L_{GaAs}\sqrt{12.89})/L_\mathrm{cav}$ (we neglect reflections at the interface between the isotropic and gyrotropic medium), $r_s$ is the Fresnel reflection coefficient at normal incidence and $D_s = 1 - r_s^2 \mathrm{e}^{2ikL_\mathrm{cav}}$. Note that since the cavity is isotropic in the $xy$-plane, we find $G_{xx}(\mathbf{k}_\parallel = 0) = G_{yy}(\mathbf{k}_\parallel = 0)$ such that $\mathbf{J}$ and consequently $\mathbf{W}$ are independent of $\sigma$. This can be used to write $H_\mathrm{int}$ in terms of the left- and right-circularly polarized modes $C_{\pm,j}(\omega)$:

$$\begin{aligned} H_\mathrm{int} &= \int_0^\infty \mathrm{d}\,\omega \sum_{i,j=1}^{n_{QW}} W_{ij}\,(\omega) \\ &\times \left\{b_i[C_{+,j}(\omega) + C_{-,j}^\dagger(\omega) + b_i^\dagger[C_{+,j}^\dagger(\omega) + C_{-,j}(\omega)]\right\}. \end{aligned} \tag{36}$$

$J_{ii}(\omega)$ is shown for a quantum well position in the center of the cavity in Fig. 1 of the main text.

The modes of the Fabry-Perot cavity can be identified as Lorentzian peaks in the spectral density, whose peak position gives the frequency of the mode, its width is related to its decay rate, and the area under the peak is equal to the square of the coupling strength (Rabi frequency)[11] . This allows us to re-derive the Hopfield model presented in the last section using our more general formalism. We find good agreement for all modes $n > 0$, but now are also able to find the decay rates of all modes and, more importantly, also the frequency and coupling strength of the $0^{\text{th}}$-order mode. Note that as we are here working in the multipolar coupling scheme, one has to convert the Rabi frequencies to the Rabi frequencies in the minimal coupling Hamiltonian employed in the last section by a factor of $\omega_c/\omega_{\text{cav}}^n$. We find $\omega_{\text{cav}}^0 = 0.058$ THz and $\Omega_0 = \sqrt{\int_0^{\omega_{max}} \mathrm{d}\,\omega J_{ii}(\omega)} = 0.47$ THz for $\omega_{\text{max}} = 0.19$ THz and on resonance, i.e. $\omega_c = \omega_{\text{cav}}^0$. Here, $\omega_{\text{max}}$ has to be chosen so that the integral covers the low-frequency peak. The value of $\Omega_0$ only slightly varies between 0.44 THz and 0.49 THz if we choose $\omega_{\text{max}} = 0.16$ or $\omega_{\text{max}} = 0.21$ THz, respectively. Moreover, it agrees with the value of $\Omega_0$ on resonance found in Eq. 18 from the Hopfield model if we consider an approximately flat normalized mode shape $f_0(z) = 1/\sqrt{L_{\text{cav}}}$, in which case we find $\Omega_0 = 0.49$ THz on resonance.

Note that the $0^{\text{th}}$ order mode has no strictly Lorentzian shape, so it would be modeled more accurately by a set of coupled photonic modes [9,10]. Such a coupled-multimode treatment is beyond the scope of the current work, as the single-mode approach taken here already provides a first approximation that is in good agreement with the measured transmission and reflection spectra, as discussed in the main text.